\documentclass[aps,pra,superscriptaddress,twocolumn]{revtex4-1}
\usepackage{dcolumn}
\usepackage[utf8]{inputenc}
\usepackage{amsmath}
\usepackage{amsfonts}
\usepackage{amssymb}
\usepackage{graphicx}
\usepackage{hyperref}
\usepackage{xcolor}

\newcommand{\ket}[1]{\ensuremath{\left|{#1}\right\rangle}}
\newcommand{\bra}[1]{\ensuremath{\left\langle{#1}\right |}}

\newcommand{\oper}[1]{\mathbf{\mathsf{#1}}}

\begin{document}

\title{Interferometric sensing of the tilt angle of a Gaussian beam}
\author{S. P. Walborn}
\affiliation{Instituto de F\'{\i}sica, Universidade Federal do Rio de Janeiro, Caixa Postal 68528, Rio de Janeiro, RJ 21941-972, Brazil}
\affiliation{Departamento de F\'{\i}sica, Universidad de Concepci\'on, 160-C Concepci\'on, Chile}
\affiliation{Millennium Institute for Research in Optics, Universidad de Concepci\'on, 160-C Concepci\'on, Chile}
\author{G. H. Aguilar}
\affiliation{Instituto de F\'{\i}sica, Universidade Federal do Rio de Janeiro, Caixa Postal 68528, Rio de Janeiro, RJ 21941-972, Brazil}
\author{P. L. Saldanha}
\affiliation{Departamento de F\'{\i}sica, Universidade Federal de Minas Gerais, 30161-970 Belo Horizonte-MG, Brazil}
\author{L. Davidovich}
\affiliation{Instituto de F\'{\i}sica, Universidade Federal do Rio de Janeiro, Caixa Postal 68528, Rio de Janeiro, RJ 21941-972, Brazil}
\affiliation{Hagler Institute for Advanced Study and Department of Physics and Astronomy, Texas A\&M University, College Station, Texas 77843, USA}
\author{R. L. de Matos Filho}
\affiliation{Instituto de F\'{\i}sica, Universidade Federal do Rio de Janeiro, Caixa Postal 68528, Rio de Janeiro, RJ 21941-972, Brazil}

\begin{abstract}
We  investigate  interferometric  techniques to estimate the deflection angle of an optical beam and compare them to the direct detection of the beam deflection.  We show that quantum metrology methods lead to a unifying treatment for both single photons and classical fields. Using the Fisher information to assess the precision limits of the interferometric schemes, we show that  the precision can be increased by exploiting the initial transverse displacement of the beam.  {This gain, which is present for both Sagnac and Mach-Zehnder-like configurations, can be considerable when compared to non-interferometric methods.}   In addition to the fundamental increase in precision, the interferometric schemes have the technical advantage that (i) the precision limits can be saturated by a sole polarization measurement on the field, and that (ii) the detection system can be placed at any longitudinal position along the beam.    We also consider position-dependent polarization measurements, and show that in this case the precision increases with the propagation distance, as well as the initial transverse displacement.    
\end{abstract}

\pacs{05.45.Yv, 03.75.Lm, 42.65.Tg}
\maketitle

\section{Introduction}
\begin{figure}
\includegraphics[width=7cm]{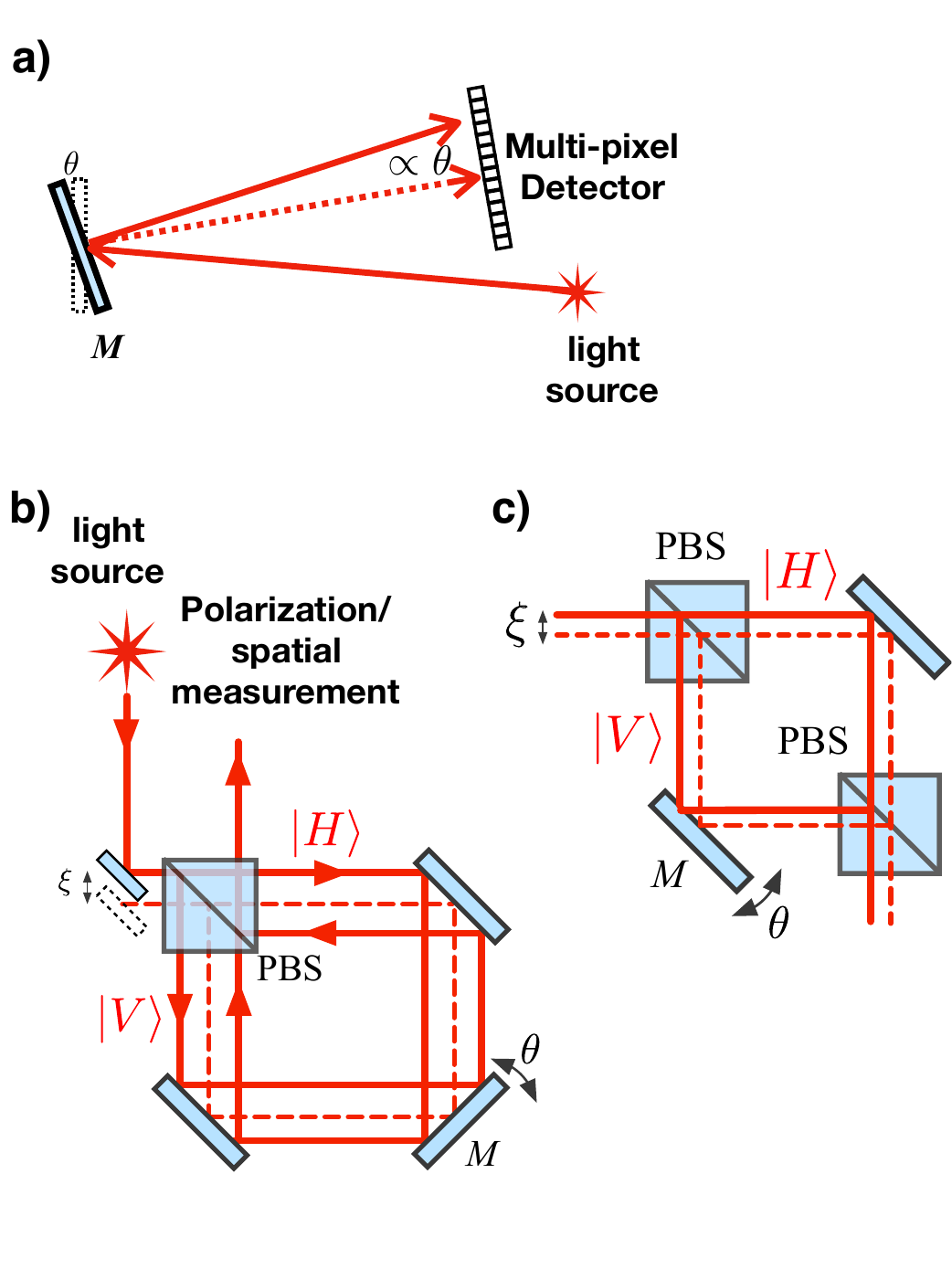}
\caption{a) Basic spatial displacement scheme, in which the transverse spatial displacement of an optical field is measured to determine the angle $\theta$. b) Sagnac interferometric method, in which the object is placed inside a Sagnac interferometer, here constructed with a polarizing beam splitter (PBS).  The counter-propagating horizontal and vertical polarization components are displaced in opposite transverse directions.  Information about $\theta$ can be obtained via spatial measurements, polarization measurements, or both. {c) Interferometric method in which the object is placed inside a Mach-Zehnder interferometer that is also constructed with PBS.} } 
\label{fig:example}
\end{figure}

There are many situations in science and tecnology that require the sensing of the mechanical motion of an object \cite{kara1981,gillies,virgo,atom}.  Several interesting and relevant techniques employ a light source to probe the object.  Consider a reflecting object that suffers a tilt of angle $\theta$.  In this case, a beam reflected from the object suffers a deviation in the reflection angle of $2 \theta$, as illustrated in Fig. \ref{fig:example}a).  By measuring the displacement of the mean position of the light beam in the far-field, one can discover $\theta$.    
\par
Several interferometric methods have been proposed to measure the deviation angle $\theta$ of an optical beam.  Some of these are based on the idea of weak-value amplification \cite{aharanov88,ribeiro08,dixon09,starling09,dressel14,alves2014weak,alves2017,martinez17}, and fall within a more general framework of post-selection based metrology \cite{tanaka13,combes14,zhang15}.  Within this framework, an interferometer is used to create patterns of constructive and destructive interference of the deviated beam. By selecting only light that appears in the appropriate destructive interference region, chosen by the physical parameters of the interferometer, an amplification of the beam displacement can be observed.  In some cases this amplification can be of several orders of magnitude \cite{dixon09}. However, the destructive interference contains very little light intensity.  Thus, there can be a loss in precision due to decreased measurement statistics.  Still weak measurements can be useful when the source intensity is much larger than the saturation threshold of the detectors, and also provides some advantages in the presence of some types of noise \cite{jordan14,viza15} and detector saturation \cite{harris17}.   
\par
Recently, it was shown that nearly optimal interferometric schemes can be achieved by considering the information gained from the post-selection statistics as well as the beam deviation \cite{alves2014weak}.  In some cases, precision bounds can be saturated by observing the post-selection statistics alone \cite{alves2017,walborn18}.  That is, by simply measuring the difference in light intensity between the bright (constructive) and dark (destructive) ports of the interferometer, an optimal amount of information can be obtained about $\theta$.  This method is thus quite simple in that it calls for the measurement of a single Stokes parameter of the light beam, and does not require a detection system with any spatial resolution.  

Furthermore, it was shown in Ref. \cite{walborn18} that polarization measurements on hyperentangled states can lead to the Heisenberg precision limit, with no need for post-selection, and that one can exploit the initial transverse displacement of the beam to extract additional information about $\theta$.   More precisely, one can achieve a precision that scales as {second moment of the relevant transverse spatial variable} $\langle x^2\rangle^{-1/2}$ as opposed to {the transverse spatial variance} $\langle \Delta_x^2 \rangle^{-1/2}$, which can be a considerable improvement.         
\par
Here we investigate interferometric-based techniques to measure the deflection angle of an optical beam.  In section \ref{sec:limits} we briefly review the relevant precision limits of optical metrology that will be employed throughout.  We employ well-known results and concepts from quantum metrology to calculate the ultimate precision limits for different experimental setups.  Section \ref{sec:dmbd} analyzes the precision achievable for the usual non-interferometric techniques based on direct measurement of the beam deflection angle using a position-sensitive detector, such as a camera, or a quadrant detector.   In section \ref{sec:IT} we introduce a simple formalism to describe interferometric techniques, where it is useful to describe the two interfering paths as an auxiliary qubit. We present both the ideal precision limits for interferometric techniques, as well as the real precision limits achievable for simple measurements on the qubit alone, which are shown to saturate the precision limits for small values of $\theta$. In addition, we consider position-dependent measurements on the qubit, and derive the available information when post-selection occurs on the spatial profile of the beam.         
%%%%%%%%%%%%%%%%%%%
%%%%%%%%%%%%%%%%%%%
\section{Precision Limits}
\label{sec:limits}
The tilt of the mirror, displayed in Fig.~\ref{fig:example}, leads to changes in the propagation of the electromagnetic field, which can be described using the tools of classical optics, if the initial field is classical. The same description would also apply to the derivation of intensity-dependent precision limits in the estimation of the tilting angle. This is not  the case, necessarily, if the initial state of the field is non-classical. A unifying treatment, encompassing both situations, is provided by quantum metrology. It can be used even for intense laser fields, for which the ultimate precision of estimation of $\theta$ can be obtained from the quantum limits for single photons, considering that the measurement of intensity $I$ of the output light is equivalent to the result of $N=I/\hbar\omega$ repeated measurements on single photons, where $\omega$ is the angular frequency of the light.  

Within the framework of quantum metrology, bounds on the uncertainty $\delta \theta$ in the estimation of a parameter $\theta$  can be assessed by the Cram\'er-Rao inequality \cite{cramer}:
\begin{equation}\label{cr1}
\delta \theta \ge \frac{1}{\sqrt{\nu F(\theta)}}\,,
\end{equation} 
where 
\begin{equation}
F(\theta)=\sum_j\frac{1}{p_j(\theta)}\left[\frac{dp_j(\theta)}{d \theta}\right]^2
\label{eq:Fisher}
\end{equation}
 is the Fisher information on $\theta$, for a given measurement on the probe used to estimate the parameter.  In the present situation, the probe is a single photon.  Here $p_j(\theta)$ is the probability of obtaining experimental result $j$, given that the value of the parameter is $\theta$, and $\nu$ is the number of repetitions of the measurement. Maximization of $F(\theta)$ over all possible quantum measurements yields the Quantum Fisher information (QFI)
${\cal F}(\theta)$, which leads to the ultimate precision bound. Since inequality \eqref{cr1} can be saturated in the limit of large $\nu$,  the Fisher information can be considered as a figure of merit for the precision of a given measurement strategy.  
\par
 For pure initial states  under unitary evolution $\oper{U}=\exp(i \theta \oper{H})$, the QFI can be calculated  by \cite{helstrom76}
\begin{equation}
\mathcal{F}(\theta) = 4 \langle \Delta \oper{H}^2 \rangle,
\label{eq:QFI}
\end{equation}
where $\langle \Delta \oper{H}^2 \rangle$ is variance of the generator of the evolution on the initial state of the probe.  From the discussion above, it is clear that the precision limit stems from the statistics of photon detections.

\section{Direct Measurement of Beam Deflection}
\label{sec:dmbd}
We refer to the ``beam deflection" technique as the direct measurement of the deflection angle of an optical beam, reflected from object $M$, as a means to infer the tilt angle $\theta$, exemplified in Fig.~\ref{fig:example}~a). We consider a Gaussian beam $G(x,y,z)$ propagating in the $z$ direction, and moreover that the beam can be separated into $x$ and $y$ components: $G(x,y,z)=E(x,z)E(y,z)$, where
\begin{equation}
E(r,z) = \left(\frac{2}{\pi w(z)^2}\right)^{\frac{1}{4}}e^{- (r-r_0)^2/w(z)^2} e^{-i kr^2/2R(z)+i\eta(z)}
\label{eq:gaussian}
\end{equation}
for $r=x,y$. Here the beam is centered at $r_0=(\xi,\zeta)$, $k$ is the wavenumber, $w(z)=w_0\sqrt{1+z^2/z_R^2}$ is the beam width at $z$, with $w_0$ being the beam waist, $R(z)=z(1+z_R^2/z^2)$ is the radius of curvature of the beam wavefront at $z$,   with $z_R=k w_0^2/2$ being the Rayleigh range --  for which the width of the beam becomes $\omega_0\sqrt{2}$ --, and $\eta(z)$ is the Gouy phase, an extra phase acquired by a propagating Gaussian beam, given by $\arctan{\left(z/z_R\right)}$ \cite{saleh91}.
We consider that the beam is incident on the reflecting object $M$ that is located at $z=0$.  When the object tilts an angle $ \theta$, the beam suffers an angular deflection $2 \theta$.  Without loss of generality, we assume that the deflection is in the $x$ direction.  This allows us to ignore the $y$ component of the optical field.  

In order to describe the beam deflection at the single-photon level,  we define the unitary operator  $\oper{U}=\exp{(-2ik\theta \oper{x})}$, where $\oper{x}$ is the transverse position operator at the object.   To quantify the ultimate precision, we use the Quantum Fisher Information \eqref{eq:QFI}.  In this case, for a single photon we have 
\begin{equation}
\mathcal{F}_{bd} = 16  k^2 \langle \Delta_x^2 \rangle, 
\label{eq:QFIbd}
\end{equation}
where $ \langle \Delta_x^2 \rangle$ is the transverse spatial variance of the beam at the object and $k$ is the wavenumber. 

For a laser field, the corresponding uncertainty is obtained via \eqref{cr1} by setting $\nu=I/\hbar\omega$, where $I$ is the total light intensity used in the measurement. 

\subsection{Position measurement}
Consider a detection system, capable of imaging the transverse profile of the laser beam.  We can calculate the Fisher information by extending \eqref{eq:Fisher} to the case of continuous measurement, resulting in
  \begin{equation}
F(\theta)=\int dx \frac{1}{P(x,\theta)}\left[\frac{d P(x,\theta)}{d \theta}\right]^2. 
\label{eq:Fishercont}
\end{equation}
For the Gaussian beam \eqref{eq:gaussian} undergoing deflection of angle $2 \theta$, the probability of detecting a photon at position $(x,z)$ is 
  \begin{equation}
  P(x,\theta,z) = A \exp\left [ -2 \frac{(x-\xi-2\theta z)^2}{w^2(z)} \right], 
\label{eq:Ppos}
\end{equation}
where $A=\sqrt{2/\pi \omega^2(z)}.$
A straightforward calculation leads to 
\begin{equation}
 F_{pos} =  16 k^2 \left (\frac{w_0^2}{4} \right) \frac{z^2}{z^2 + z_R^2}.
\label{eq:pos}
\end{equation}
Considering that the term in parentheses is the variance of the transverse spatial distribution of the Gaussian beam in Eq. \eqref{eq:gaussian}, this shows that a high-resolution position measurement saturates the QFI \eqref{eq:QFIbd} when $z >> z_R$.  Moreover, the transverse displacement of the beam, $\xi$, plays no role in the Fisher information for this specific case.  
\subsection{Quadrant measurement}
The mean position of a Gaussian beam can be measured using a quadrant detector (here used as a ``sign" detector) to measure the signal that is proportional to the beam intensity on the positive ($+$) or negative side ($-$) of the $x$-axis. {For a quadrant detector at position $z$ and origin set at $x=\xi$, this can be characterized by two probabilities $P_\pm$, obtained by integrating probability \eqref{eq:Ppos} over positive or negative $x$, giving}
\begin{equation}
 P_{\pm} =  \frac{1}{2}\left[ 1  \pm \mathrm{erf}\left(\frac{2 \sqrt{2} \theta z}{w(z)} \right) \right ].
\label{eq:Pquad}
\end{equation}
Then, using definition \eqref{eq:Fisher}, the quadrant detection system gives a Fisher information
\begin{equation}
 F_{quad} =  \frac{32 z^2 e^{-2\left(\frac{2 \sqrt{2} \theta z}{w(z)} \right)^2}}{\pi w^2(z)\left[ 1  - \mathrm{erf}^2\left(\frac{2 \sqrt{2} \theta z}{w(z)} \right) \right ]}.
\label{eq:Fquad}
\end{equation}
For small tilt angles, so that $\left(\frac{2 \sqrt{2} \theta z}{w(z)} \right)^2 <<1$, we have 
\begin{equation}
 F_{quad} =  \frac{32 z^2 }{\pi w^2(z)}. 
\label{eq:Fquad2}
\end{equation}
 Note that $z^2/w^2(z)$ varies from $0$ when $z=0$ to $z_R^2/w_0^2$ when $z$ is much larger than the Rayleigh range  $z_R$.  Thus, using $z_R=\pi w_0^2/\lambda$, {we can write the maximum Fisher information for the quadrant detector as}
 \begin{equation}
 F_{quad,max} =  \frac{32 z_R^2 }{\pi w^2_0} =   \frac{32}{\pi} k^2 \left( \frac{w_0^2}{4} \right),
\label{eq:Fquad3}
\end{equation}
which is achieved for $z >> z_R$, where the detected field is the Fourier transform of the field at the reflecting object.  Note that  the term in parentheses on the right is again the beam variance at $z=0$.  This expression is $2/\pi \approx 0.64$ times the QFI given in \eqref{eq:QFIbd}, and thus does not extract all of the available information on $\theta$.  
\section{Interferometric Techniques}
\label{sec:IT}
\subsection{Quantum Fisher Information - Ideal Precision Limits}
Let us now consider interferometric schemes to optically measure the tilt angle $\theta$ of a reflecting object $M$, again located at longitudinal position $z=0$. We consider first a Sagnac interferometer, as shown in Fig.~\ref{fig:example} b).  Considering that interferometer is constructed with a PBS, the two optical paths in the interferometer are associated to the horizontal $H$ and vertical $V$ polarization directions.  Thus, the polarization of the beam acts as an auxiliary qubit.  The two orthogonally polarized fields propagate in opposite directions within the interferometer.     
The unitary operator describing the dynamics of this interferometric scheme can be written as \cite{alves2014weak,alves2017,walborn18}
\begin{equation}
\oper{U}_{sag}=\exp{(-2ik\theta \sigma_z \oper{x})},
\label{eq:Usag}
\end{equation}
where $\sigma_z = \ket{H}\bra{H}- \ket{V}\bra{V}$ is the Pauli operator acting on the auxiliary (polarization) degree of freedom.  Assuming that the initial polarization state is isotropic, so that $\langle \sigma_z \oper{x}\rangle = \langle \sigma_z\rangle \langle \oper{x}\rangle $, the Quantum Fisher Information is given by 
\begin{equation}
\mathcal{F}_{sag} = 16 k^2 [\langle \Delta_x^2 \rangle + (1-\langle \sigma_z \rangle^2)\langle \oper{x}\rangle^2],
\label{eq:QFIint}
\end{equation}
 so that the initial displacement of the field $\langle \oper{x}\rangle$ can be used to increase the measurement precision \cite{walborn18}, when compared to the direct measurement of beam deflection, in which the QFI is given by  Eq. \eqref{eq:QFIbd}.  Of course, by choosing the polarization state such that $\langle \sigma_z \rangle=\pm1$, the interferometric scheme is equivalent to the displacement scheme (no interference).  However, one can increase the measurement precision by taking advantage of the coherence in the additional (polarization) degree of freedom, choosing an initial state with $\langle \sigma_z \rangle=0$ and $\langle \oper{x} \rangle \neq 0$.  These precision bounds can be saturated by using practical setups.  Depending on the initial polarization state, the relevant information on $\theta$, can be concentrated in the polarization or spatial degrees of freedom, or both \cite{alves2014weak,alves2017,walborn18}. 
 \par
 {One might consider other types of optical interferometers, such as a Mach-Zehnder interferometer, as shown in Fig. \ref{fig:example} c), where again we associate the two paths with $H$ and $V$ polarization directions. Here only one path--say the $V$ polarized on -- probes the tilting object.   This would also be the case for a Michelson interferometer, for which a similar analysis applies.  The unitary operator describing these dynamics can be written as  
\begin{equation}
\oper{U}_{mz}=\exp{(-2ik\theta \ket{V}\bra{V}  \oper{x})}. 
\end{equation}
To calculate the Quantum Fisher Information, we assume again an isotropic polarization state, and find
\begin{equation}
\mathcal{F}_{mz} = 8 k^2 (1-\langle \sigma_z \rangle) [\langle \Delta_x^2 \rangle + \frac{1}{2}(1+\langle \sigma_z \rangle)\langle \oper{x}\rangle^2]. 
\label{eq:QFmz}
\end{equation}
If the initial polarization state is $V$, so that $\langle \sigma_z \rangle=-1$, we recover the QFI for the beam displacement method \eqref{eq:QFIbd}.  Of course if $\langle \sigma_z \rangle=1$ ($H$ polarization) the QFI is zero, since the beam does not reflect off the object, and no information is obtained.  In general the optimal polarization state depends on the ratio between the variance and the squared mean position of the input beam.  If $\langle \oper{x}\rangle^2 >> \langle \Delta_x^2 \rangle$, it is optimal to choose  a polarization state such that $\langle \sigma_z \rangle=0$, and we have  $\mathcal{F}_{mz} = 8 k^2  [\langle \Delta_x^2 \rangle + \langle \oper{x}\rangle^2/2]$. In fact, we can exploit the interference (choosing $\langle \sigma_z \rangle=0$) to surpass  the QFI for the beam displacement method \eqref{eq:QFIbd} when the mean is more than twice the variance.  Moreover, in this case it is enough to measure only the polarization of the beam, as will be discussed more detail in the following section.  For both the Mach Zehnder and Sagnac interferometer, the increase in the QFI as a function of $\langle x \rangle^2$ is due to an extra phase that results from the fact that the light hits  $M$ out of center, as illustrated in Fig. \ref{fig:mirrorspecs}. For a beam with a given variance and mean, we have $\mathcal{F}_{mz} < \mathcal{F}_{sag}$.  This is of course due to the fact that the Sagnac interferometer allows both counter-propagating beams to probe the angle $\theta$, which increases the attainable precision.      
 }
\subsection{Precision for practical interferometric scheme}
{Since the QFI for the Sagnac configuration of Fig.~\ref{fig:example} b) is larger than that of the Mach-Zehnder configuration, we consider a Sagnac interferometer in what follows, where we describe the polarization and transverse spatial degrees of freedom of the light beam using Dirac notation \cite{stoler81,tasca11}.}
We choose the initial polarization/spatial state as 
\begin{equation}
\ket{\Psi}=(\alpha \ket{H} + \beta \ket{V}) \ket{\psi},
\end{equation} 
where $\alpha$ and $\beta$ are complex coefficients,
\begin{equation}
\langle q \ket{\psi} = \int dq \psi (q)e^{-i q \xi}, 
\end{equation}
and $\ket{q}$ is a plane wave with transverse component $q$ in the $x$ direction, and wavenumber $k$.  Here we assume that the initial displacement of the beam is $\langle x \rangle = \xi$.  

Applying the unitary interaction for the Sagnac interferometer \eqref{eq:Usag}, together with free propagation, described by operator $\exp{(i \oper{q}^2z/2k})$ \cite{stoler81,tasca11}, we have
\begin{equation}
\ket{\Psi}=  \alpha \ket{H}\ket{\psi_+}+ \beta \ket{V}\ket{\psi_-}
\end{equation}
where
\begin{equation}
\ket{\psi_\pm}= \int dq \psi (q) e^{-i q \xi} e^{i (q\pm2k\theta)^2z/2k}\ket{q\pm2k\theta}. 
\end{equation}
Consider that the spatial state $\psi(q)$ describes the Gaussian beam \eqref{eq:gaussian} in the momentum representation.  We can calculate the detection amplitude at position $x$ by projecting the state onto $\ket{x}$.  We assume also that projection onto diagonal linear polarization states$\ket{\pm}=(1/\sqrt{2})(\ket{H}\pm \ket{V})$, has been performed.   The probability $P_{\pm}(x)$ to detect a photon at position $x$ in the $\ket{\pm}$ polarization state is 

\begin{align}
\label{eq:P}
P_{\pm}(x) = & \frac{|\alpha|^2}{2}|\psi(x-\xi+2\theta z,z)|^2 + \frac{|\beta|^2}{2}|\psi(x-\xi-2\theta z,z)|^2  \\ \nonumber 
&  \pm A d e^{\frac{-8\theta^2z^2}{w^2(z)}}e^{\frac{-2(x-\xi)^2}{w^2(z)}} \times \\ \nonumber 
&  \cos\left(\frac{4 k  \theta w_0^2 }{w(z)^2}(x-\xi) +  {4 k \theta \xi} - \varphi\right),
\end{align}
where we define $d \exp(i \varphi) = \alpha^*\beta$, with $d$ a real number. 
\par
\subsubsection{Polarization measurement}
Let's first consider polarization measurement alone.  Integrating \eqref{eq:P} over $x$ and using the relations between the Gaussian beam parameters, we have
\begin{equation}
 P_{\pm} =  \frac{1}{2}\left [ 1  \pm 2 d\,  e^{-B \theta^2}\cos(4 k \theta \xi - \varphi) \right],
\label{eq:Pint}
\end{equation}
where $B=2 k^2 w_0^2$. 
We note that these two probabilities can be used to determine the Stokes parameter of the field: $S_z=P_+ - P_-$.  
The Fisher information is 
\begin{equation}
F =  \frac{16 d^2[B \theta \cos (4 k \xi \theta- \varphi)+2 k \xi \sin (4 k \xi \theta- \varphi)]^2}{e^{2B\theta^2}-4 d \cos^2 (4 k \xi \theta- \varphi)}.
\label{eq:Fpolonly}
\end{equation}
{We note that $F$ depends on the initial polarization state via $d$ and $\varphi$.  Using a Bloch sphere representation for the polarization qubit, without loss of generality we can write $\alpha=\cos(\vartheta/2)$ and $\beta=\exp(i \varphi) \sin(\vartheta/2)$, where $\vartheta$ and $\varphi$ are the angular coordinates in the unit sphere \cite{chuang00}. With this,  $2 d = \sin\vartheta$.  To maximize $F$ for any value of $\xi$, we must maximize $d$.  To achieve $\sin \vartheta = 1$ we require an initial polarization state that lies on the equator of the Bloch sphere, for which we achieve maximum contrast interference, since these have equal projections onto the $H$ and $V$ polarization states.  The relative phase $\varphi$ between $H$ and $V$ polarization components of the initial state can be used as a control parameter to maximize $F$ for a particular range $\vartheta$.  We note that a very similar analysis holds if one fixes the input state and maximizes over the measurements.} 

Thus, to maximize the Fisher information \eqref{eq:Fpolonly}, we set $\alpha=\beta=1/\sqrt{2}$, corresponding to the initial polarization state $\ket{+}$.
Assuming $B\theta^2 <<1$ and $4 k \xi \theta <<1$, and keeping only terms up to first order in $\theta$, the above expression reduces to 
\begin{equation}
F = 16 k^2\left [\frac{w_0^2}{4} + \xi^2 \right ].
\label{eq:FishPol2}
\end{equation}
Noting that $\langle \Delta^2 x \rangle = w^2_0/4$, the Fisher information \eqref{eq:FishPol2} is equal to the QFI \eqref{eq:QFIint}, and superior to the QFI for the beam displacement technique \eqref{eq:QFIbd}, due to the initial translation of the input beam $\langle x \rangle = \xi$.  
\par
\begin{figure}
\includegraphics[width=6cm]{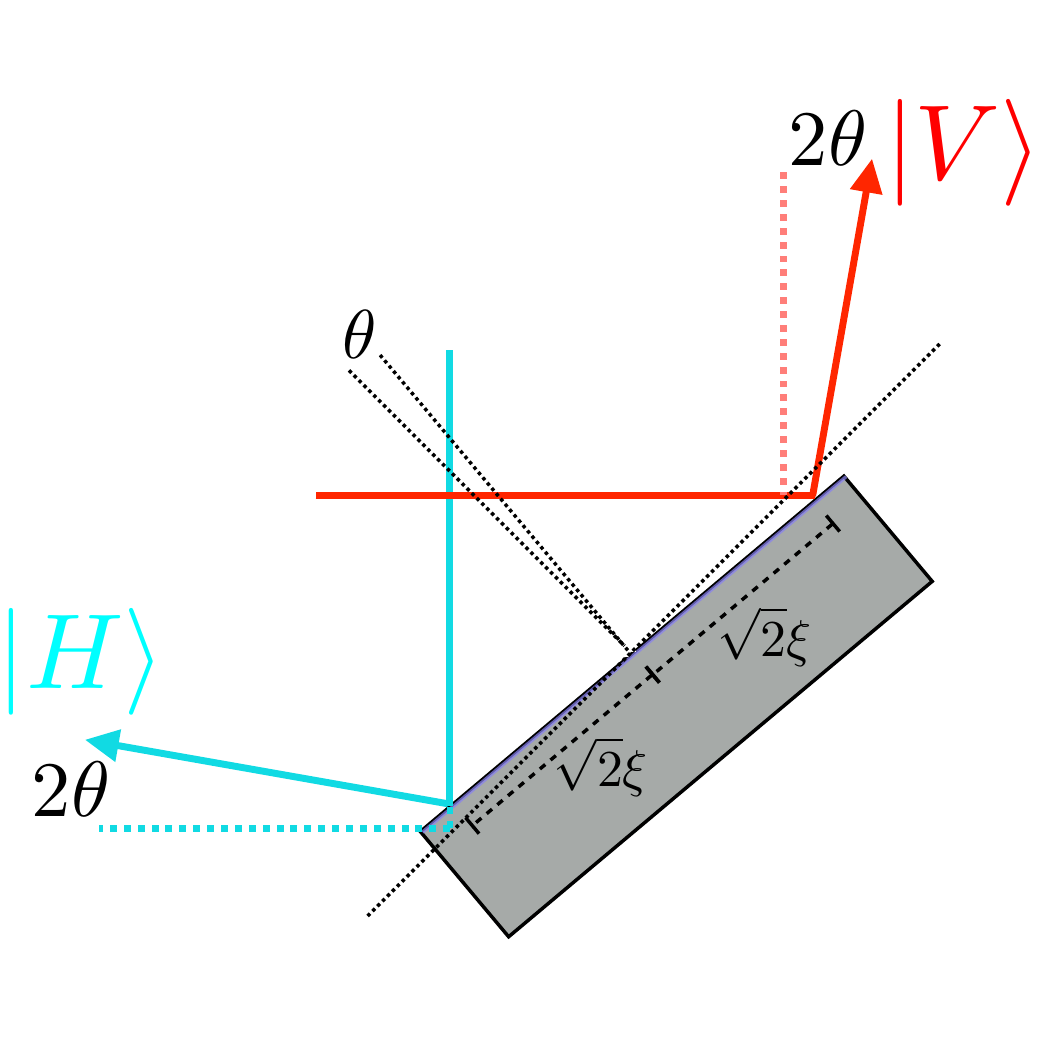}
\caption{{Diagram showing the origin of information gain due to transverse beam displacement $\xi$ of the inital beam in the Sagnac interferometer.  The two counter-propagating paths are displaced $\sqrt{2} \xi$ on the surface of the reflecting object $M$.  When the surface tilts by an angle $\theta$ relative to the initial position (dotted black line at 45$^\circ$), one beam (the $V$ beam in this example) propagates a distance $2 \theta \xi$ longer to reach the surface, while the other propagates a distance $2 \theta \xi$ shorter. } }  
\label{fig:mirrorspecs}
\end{figure}

{To explain the physical origin of this gain in information on $\theta$ due to $\xi$, we refer to Fig.~\ref{fig:mirrorspecs}, which shows a diagram of the reflecting object $M$ with two counter-propagating beams with $H$ and $V$ polarization.  When the surface tilts by an angle $\theta$ relative to its initial position (black dotted diagonal line), the $V$ beam propagates a distance $2 \theta \xi$ longer to reach the surface, while the other propagates a distance $2 \theta \xi$ shorter.  This causes a relative phase difference of $4 \theta \xi$ between the two orthogonal beams, creating a change of the polarization state that contains information on $\theta$, and depends also on $\xi$.}
We note that this precision is achieved by polarization measurements alone, with no spatial resolution required in the detection system.  Moreover, the interferometric technique has the advantage over the beam deflection method in that the polarization measurements can saturate the QFI \eqref{eq:QFIint} at nearly any propagation distance $z$, as opposed to being restricted to the far-field ($z>>z_R$). 

%%%%%%%%%%%%%%%%%%%
%%%%%%%%%%%%%%%%%%%
\par

\subsubsection{Position-dependent polarization measurement}
Several authors have considered post-selected metrology schemes \cite{tanaka13,combes14,zhang15}.  However, most previous work has considered a position measurement implemented with post-selection on the auxiliary qubit (polarization) \cite{ribeiro08,dixon09,starling09,tanaka13, alves2014weak,alves2017,zhang15}. The fact that we can extract more information from the polarization degree of freedom in the interferometric scheme motivates the consideration of polarization measurements conditioned on position post-selection.   We will see that this analysis reveals the origin of the information on $\theta$ at different propagation distances $z$. 
\par
Let us consider that point-like detectors are placed at position $x$, so that a position-dependent projection onto the polarization states $\ket{\pm}$ is performed. 
We can define normalized probabilities for the polarization measurements, 
\begin{equation}
\bar{P}_\pm(x) = \frac{{P}_\pm(x)}{{P}_+(x)+{P}_-(x)}
\end{equation}
such that $\bar{P}_+(x) + \bar{P}_{-}(x)=1$.  The Fisher information regarding polarization measurement at point $x$ can then be written as
\begin{equation}
\bar{F}(x) = \frac{1}{\bar{P}_+(x)\bar{P}_-(x)}\left(\frac{d \bar{P}_+(x)}{d \theta} \right)^2.
\end{equation}
Following \cite{combes14,alves2014weak}, the total Fisher information for both position and polarization measurements is then given by 
\begin{equation}
{F} = \int dx P(x) \bar{F}(x) + \int dx \frac{1}{{P}(x)}\left(\frac{d {P}(x)}{d \theta} \right)^2,
\end{equation}
where $P(x)=P_{+}(x)+P_{-}(x)$.  The first term is the average Fisher information obtained for position-dependent polarization measurements, while we can recognize the second term as the Fisher information from the position measurements.   We note that  $\bar{F}(x)$ can thus be interpreted as the Fisher information per photon detected at $x$, with $P(x)$ being the probability that a photon is detected at $x$. 
\par
 Let us explore the information available in position-dependent polarization measurements. We assume that $\alpha=\beta=1/\sqrt{2}$ in Eq \eqref{eq:P}, as we saw above that this choice optimizes the Fisher information.  Explicitly, the normalized probabilities read
\begin{equation}
\bar{P}_{\pm}(x) = \frac{1}{2} \left (1  \pm \frac{\cos\left(\frac{4 k  \theta w_0^2 }{w(z)^2}(x-\xi)+ {4 k \theta \xi}\right)}{\cosh\left(\frac{8 \theta z(x-\xi)}{w^2(z)}\right)}\right).
\end{equation}
Writing 
\begin{equation}
\bar{P}_{\pm}(x) = \frac{1}{2} \left(1  \pm \frac{\cos [a(x) \theta]}{\cosh [b(x) \theta]} \right), 
\end{equation}
and assuming $\theta$ sufficiently small so that $ a(x)^2 \theta^2<<1$ and $b(x)^2 \theta^2 <<1$. These can be re-written as 
 \begin{equation}
 a(x)^2 \theta^2 =  16 k^2 \theta^2\left(\frac{z_R^2 x + z^2 \xi }{z^2+z_R^2} \right)^2 <<1,
 \end{equation} 
 and
\begin{equation}
b(x)^2 \theta^2  = 16 k^2 \theta^2 \left(\frac{z z_R (x-\xi)}{z^2+z_R^2}\right)^2 << 1.
\end{equation}

Let us briefly analyze these assumptions in more detail.  For measurements in the far-field, we consider $z>>z_R$. The above conditions return the constraint $16 k^2 \theta^2 \xi^2 <<1$. For near-field measurements, let us first note that throughout this section we have assumed that any measurement will be realized after the field has exited the interferometer (see Fig.~\ref{fig:example} b)), and that we have defined $z=0$ as the surface of the reflecting object, within the interferometer.  Thus, to consider measurements in the near-field, we take $z<<z_R$, with the understanding that $z$ must be larger than the propagation distance from the reflecting object to the exit face of the interferometer.  Under these conditions, and assuming finite $x$ and $\xi$, the above assumptions require  $16 k^2 \theta^2 x^2 <<1$. Since $x^2 \sim w_0^2/4$, these two assumptions place the same limits on the size of deflection angle $\theta$ relative to the initial beam displacement $\xi$ and the beam width $w_0$ as was considered just before Eq. \eqref{eq:FishPol2}.
Then, we have, up to $O(\theta^2)$
\begin{equation}
\bar{F}(x) \approx a(x)^2 + b(x)^2, 
\end{equation}
or 

\begin{equation}
\bar{F}(x) \approx 16 k^2 \frac{z_R^2 x^2+z^2\xi^2}{z^2+z_R^2}. 
\label{eq:Fx}
\end{equation}
This equation clearly displays the two contributions to the Fisher information, where one can see that the initial beam displacement $\xi$ is most relevant in the far-field ($z>>z_R$), while the beam width, through $|x| \sim w_0$, is most relevant in the near-field ($z<<z_R$).
%%%%%%%%%%%%%%%%%%%%%%%%%%%%%%
\begin{figure}
\includegraphics[width=6cm]{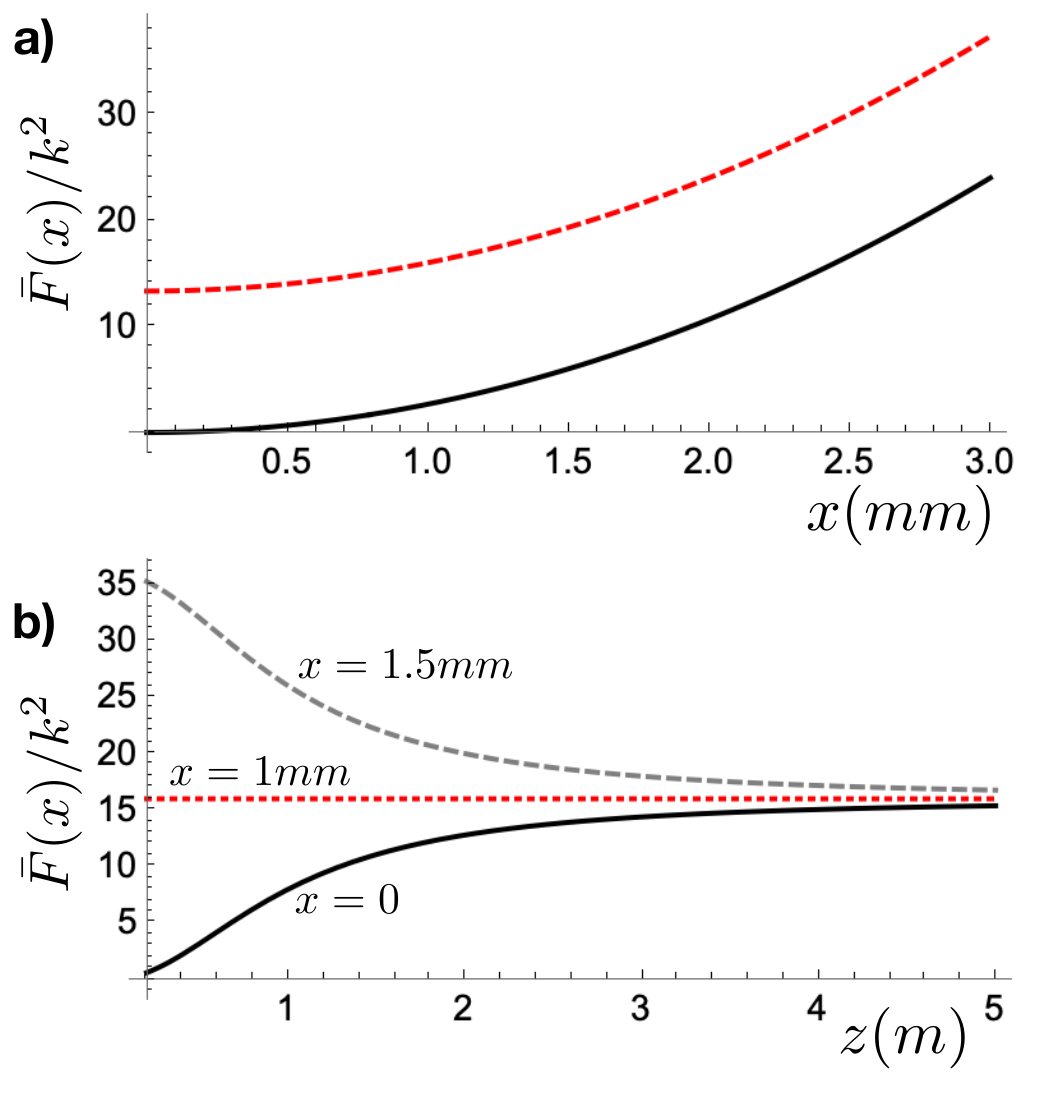}
\caption{a)  Fisher information per detected photon $\bar{F}(x)/k^2$ at $z=5 z_R$, as a function of $x$, for $x$-dependent polarization measurement.  Solid black line and dashed red line correspond respectively to displacements of the incoming beam equal to $\xi=0$ and $\xi=1$mm.  b) Fisher information per detected photon $\bar{F}(x)/k^2$ as a function of $z$ for polarization measurement  for $\xi=1$mm. The position measurements are made at $x=0$ (solid black curve), $x=1$mm (dotted red line), and $x=1.5$mm (grey dashed curve). In all plots we chose $z_R=1$m.} 
\label{fig:Fisher_Plots}
\end{figure}
%%%%%%%%%%%%%%%%%%%%%%%%%%%%%%
Figure \ref{fig:Fisher_Plots} a) shows a plot of $\bar{F}(x)/k^2$ with $\bar{F}(x)$ given by \eqref{eq:Fx}  as a function of $x$ for two different values of the displacement $\xi$.   We can see that the information gain arising from the beam displacement, namely $16k^2\xi^2$, appears for any detection position $x$.   The Fisher information increases quadratically with $x$, however these events are more and more rare, as the probability $P(x)$ decreases with $x$.  For $x<<\xi$, we have $\bar{F}_{x<<\xi}(x) \approx16 k^2 \xi^2z^2/(z^2+z_R^2)$, showing that the information in the polarization degree of freedom depends on the propagation distance $z$, saturating at $16 k^2 \xi^2$ for $z >> z_R$.  This is the information that arises from the phase shift due to the initial transverse displacement $\xi$ of the input beam. This can be seen in the plots in figure  \ref{fig:Fisher_Plots} b), showing $\bar{F}(x=0)/k^2$ as a function of $z$ for $\xi = 1mm$. We can see from Eq.~\eqref{eq:Fx}  that at $x=0$ and $z<< z_R$,  there is no information available in the polarization degree of freedom, $\bar{F}(0)=0$, for any value of $\xi$.  Thus, if the detection system has a limited spatial bandwidth accepting light near $x=0$, it is advantageous to detect light in the far-field $z >> z_R$.  On the other hand, for larger values of $x$, $\bar{F}(x)$ is maximum for small $z$ and decreases as a function of $z$.   We can see from \eqref{eq:Fx}  and Fig.~\ref{fig:Fisher_Plots} b) that the critical point is $x=\xi$, where the Fisher information is constant as a function of $z$.  
\par
\begin{figure}
\includegraphics[width=8cm]{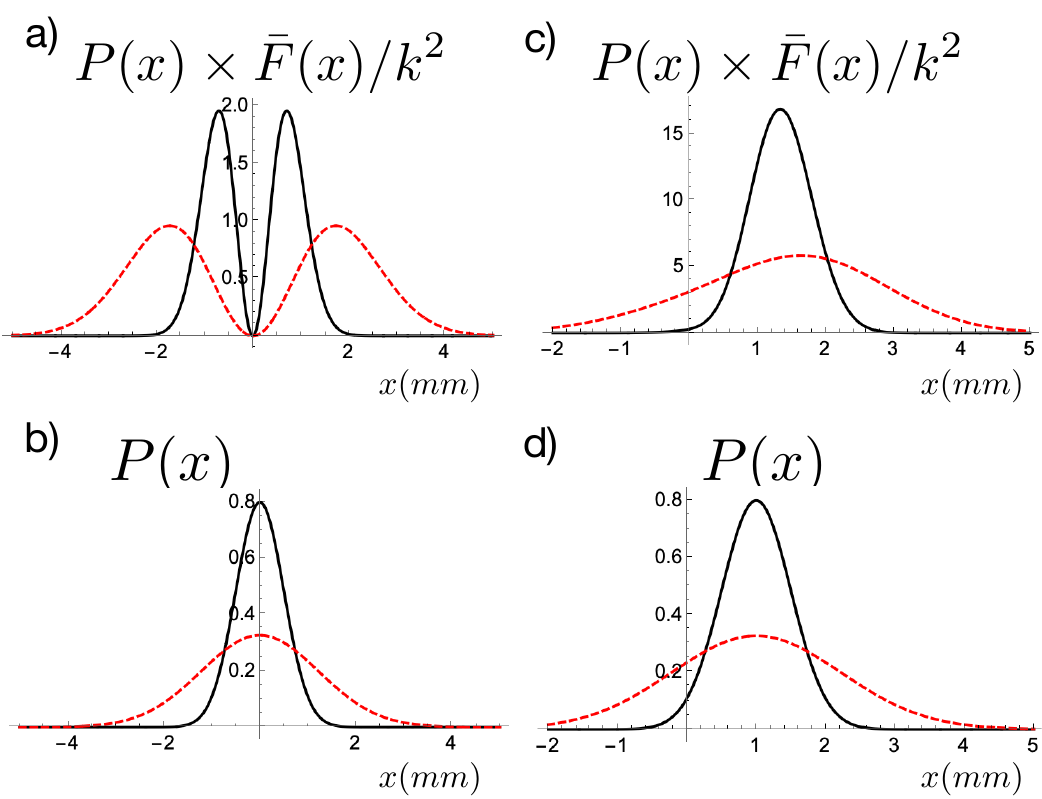}
\caption{ a) Fisher information for  $\xi=0$ scaled by the  detection probability $P(x)$ shown in b).  c) Fisher information for  $\xi=1$mm scaled by the detection probability $P(x)$ shown in d). Black curves correspond to $z=0$ and red dashed curves to $z=5 z_R$.  In all plots we chose the Rayleigh range $z_R=1$m.} 
\label{fig:Scaled_Fisher_Plots}
\end{figure}

Of course the actual information that can be extracted depends upon the probability of detecting a photon at a given point $x$.  In Fig.~\ref{fig:Scaled_Fisher_Plots} we show the Fisher information per photon at $x$ scaled by the actual probability of detecting a photon at $x$.  This quantity, $P(x) \times \bar{F}(x)/k^2$ represents the information on $\theta$ per detected photon (at $x$) obtainable by projecting the polarization in the $\ket{\pm}$ basis.  There can also be information in $P(x)$ alone. We can see that the maximum scaled Fisher information at $x$ changes as a function of $\xi$ and $z$.  Thus, the post-selection information $P(x) \bar{F}(x)$ depends upon $z$.  This is in contrast to a full-field detection system, which registers all available light and saturates the QFI for all values of $z$ (outside the interferometer).
We can check average Fisher information obtained from polarization measurements is obtained by averaging over $P(x)$,  which gives 
\begin{equation}
\int dx P(x) \bar{F}(x) =  16 k^2\left [\frac{w_0^2}{4} + \xi^2 \right],
\end{equation}
which agrees with \eqref{eq:FishPol2}, as it should.  {Since this saturates the QFI \eqref{eq:QFIint}, there is no available information in the spatial degree of freedom. }

 \section{Conclusions}
We have investigated interferometric-based techniques to estimate the deflection angle $\theta$ of an optical beam, comparing them with the direct detection of the beam deflection in the far field.  

Using quantum metrology tools, we compared the ultimate precision limits given by the Quantum Fisher Information for both a Sagnac interferometer and a Mach-Zehnder interferometer, and observed that the Sagnac offers superior precision for fixed beam parameters. For both of these arrangements,  we have shown that the ultimate precision limit can surpass that of a far-field beam displacement measurement by a factor that is proportional to the initial transverse displacement of the beam.  Thus, the initial beam displacement, a parameter independent from $\theta$, can be used to enhance the measurement precision.  

Moreover, {focusing on the Sagnac approach}, we have shown that this precision limit can be saturated for small $\theta$ by a binary polarization measurement of the light field.  An additional advantage, when compared to beam displacement methods, is that the polarization measurement can be performed at nearly any longitudinal position  after the tilt angle has been imprinted on the beam.  Thus, the detector can also be placed in the near-field of the object.   
\par We have also considered polarization measurements conditioned on post-selection of the position by point-like detectors.  This analysis also illuminated the role of the two parameters which contribute to the Fisher information: the initial beam displacement is most relevant in the far-field of the object, while the beam width is most relevant in the near-field. 
It is also relevant for situations in which the full-field measurement of the field is not possible. Near the intensity maximum at the origin, the information in the polarization degree of freedom is minimum in the near-field and increases for detection in the far-field. For an initially non-displaced beam, in the near field the information in the polarization degree of freedom is maximum for detections performed outside the beam center.  
Through the use of the tools of quantum metrology, our analysis was shown to be valid for single photons and classical fields as well, thus representing a unified treatment for both situations.  We note that a similar approach in optical imaging has led to interesting new measurement schemes \cite{tsang16,paur16}. 
\par

\begin{acknowledgements}
The authors would like to thank CAPES, CNPQ (PQ 307058/2017-4), FAPERJ (E-26/202.790/2017), and the National Institute of Science and Technology for Quantum Information (INCT-IQ) for financial support.  This work was realized as part of the CAPES/PROCAD program. SPW received support from the Fondo Nacional de Desarrollo Científico y Tecnológico (Conicyt) (1200266) and the Millennium Institute for Research in Optics. 
\end{acknowledgements}

%\bibliographystyle{apsrev}
%\bibliography{Inter-metro}

  %%%%%%%%%%%%%%%%%%%%%%%%%%%%%%%%%%%%%%%%%%%%%%%%%%%%%%%%%%

\end{document}